\documentstyle[epsfig,12pt]{article}
\newcommand{\be}{\begin{eqnarray}}
\newcommand{\ee}{\end{eqnarray}}

\def\lsim{\mathrel{\rlap{\lower4pt\hbox{\hskip1pt$\sim$}}
\raise1pt\hbox{$<$}}}               % less than or approx. symbol
\def\gsim{\mathrel{\rlap{\lower4pt\hbox{\hskip1pt$\sim$}}
\raise1pt\hbox{$>$}}}               % greater than or approx. symbol

\begin{document}

\begin{figure}[htb]

\epsfxsize=6cm \epsfig{file=logo_INFN.epsf}

\end{figure}

\vspace{-4.75cm}

\Large{\rightline{Sezione ROMA III}}
\large{
\rightline{Via della Vasca Navale 84}
\rightline{I-00146 Roma, Italy}
}

\vspace{0.6cm}

\rightline{INFN-RM3 98/5}
\rightline{October 1998}

\normalsize{}

\vspace{2cm}

\begin{center}

\Large{ANALYSIS OF THE $\Lambda_b \to \Lambda_c + \ell \bar{\nu}_{\ell}$ 
DECAY WITHIN A LIGHT-FRONT CONSTITUENT QUARK MODEL}\\

\vspace{1cm}

\large{Fabio Cardarelli and Silvano Simula\footnote{\bf To appear in 
Physical Review D.}}

\vspace{0.5cm}

\normalsize{Istituto Nazionale di Fisica Nucleare, Sezione Roma III,\\ Via
della Vasca Navale 84, I-00146 Roma, Italy}\\

\end{center}

\vspace{1cm}

\begin{abstract}

\noindent We present an investigation of the Isgur-Wise form factor relevant
for the semileptonic decay $\Lambda_b \to \Lambda_c + \ell \bar{\nu}_{\ell}$
performed within a light-front constituent quark model. Adopting different
baryon wave functions it is found that the Isgur-Wise form factor depends
sensitively on the baryon structure. It is shown however that the shape of
the Isgur-Wise function in the full recoil range relevant for the $\Lambda_b
\to \Lambda_c + \ell \bar{\nu}_{\ell}$  decay can be effectively constrained
using recent lattice $QCD$ results at low recoil. Then, the $\Lambda_b \to
\Lambda_c + \ell \bar{\nu}_{\ell}$ decay is investigated including both
radiative effects and first-order power corrections in the inverse
heavy-quark mass. Our final predictions for the exclusive semileptonic
branching ratio, the longitudinal and transverse asymmetries, and the
longitudinal to transverse decay ratio are: $Br(\Lambda_b \to \Lambda_c
\ell \bar{\nu}_{\ell}) = (6.3 \pm 1.6) ~ \% ~ |V_{bc} / 0.040|^2 ~
\tau(\Lambda_b) / (1.24 ~ ps)$, $a_L = -0.945 \pm 0.014$, $a_T = -0.62 \pm
0.09$ and $R_{L/T} = 1.57 \pm 0.15$, respectively. Moreover, both the
longitudinal asymmetry and the (partially integrated) longitudinal to
transverse decay ratio are found to be only marginally affected by the model
dependence of the Isgur-Wise form factor as well as by first-order power
corrections; therefore, their experimental determination might be a very
interesting tool for testing the $SM$ and for investigating possible New
Physics.

\end{abstract}

\newpage

\pagestyle{plain}

\section{Introduction}

\indent The Heavy Quark Effective Theory ($HQET$) is widely recognised as an
appropriate theoretical framework for analysing the properties of baryons
containing a single heavy quark ($Q$). It provides a systematic expansion of
the $QCD$ Lagrangian as a series of powers of the inverse heavy-quark mass
($m_Q$). At the leading order a new spin-flavour symmetry, named the Heavy
Quark Symmetry ($HQS$), arises. Such a symmetry is shared by, but not
manifest in $QCD$ and it is broken by radiative corrections as well as by
non-perturbative contributions, which can be organised as an expansion in
powers of $1 / m_Q$. The $HQS$ allows to derive several model-independent
relations among hadronic properties and, in particular, all the electroweak
transition and elastic amplitudes can be expressed in terms of a subset of
{\em universal} form factors \cite{iw}. In the case of the semileptonic
decay process $\Lambda_b \to \Lambda_c + \ell \bar{\nu}_{\ell}$ the vector
and axial-vector transition amplitudes, $\langle \Lambda_c(v') | \bar{c}
\gamma^{\mu} b | \Lambda_b(v)\rangle$ and $\langle \Lambda_c(v') | \bar{c}
\gamma^{\mu} \gamma^5 b | \Lambda_b(v)\rangle$, where $v$ ($v'$) is the
initial (final) baryon four-velocity, can be expressed in terms of only one
universal function \cite{iw}, called the Isgur-Wise ($IW$) form factor
$\xi(\omega)$, where $\omega \equiv v \cdot v'$. However, except for the
normalisation condition $\xi(1) = 1$, the $HQS$ cannot predict the full
behaviour of the $IW$ form factor, for the complete knowledge of the
non-perturbative baryon structure is required. Therefore, calculations based
on lattice $QCD$ simulations, effective approaches and models are necessary
in order to obtain reliable quantitative predictions, which could allow to
extract from the data important information on fundamental parameters, like,
e.g., the $CKM$ weak mixing angles, and on possible extensions of the
Standard Model ($SM$).

\indent In this work we adopt a relativistic constituent quark model
formulated on the light-front ($LF$), and, using different types of $Q q q'$
wave functions (where $q$ and $q'$ are two light spectator quarks), we
extend our investigation started in Ref. \cite{plb98} concerning the $IW$
form factor $\xi(\omega)$ relevant for the decay process $\Lambda_b \to
\Lambda_c + \ell \bar{\nu}_{\ell}$. Our aim is to constrain as much as
possible the model dependence of the calculated $IW$ form factor in order to
estimate several observables, like the exclusive semileptonic branching
ratio and various (integrated) asymmetries. To this end we will make use of
recent lattice $QCD$ simulations \cite{UKQCD} at low recoil, so that our
model can be viewed as a lattice-constrained $LF$ quark model. After
including both radiative effects and first-order $1 / m_Q$ corrections to
the relevant form factors, our final results for the exclusive semileptonic
branching ratio, the longitudinal and transverse asymmetries, and the
longitudinal to transverse decay ratio are: $Br(\Lambda_b \to \Lambda_c \ell
\bar{\nu}_{\ell}) = (6.3 \pm 1.6) ~ \% ~ |V_{bc} / 0.040|^2 ~
\tau(\Lambda_b) / (1.24 ~ ps)$, $a_L = -0.945 \pm 0.014$, $a_T = -0.62 \pm
0.09$ and $R_{L/T} = 1.57 \pm 0.15$, respectively. The theoretical
uncertainties on $Br(\Lambda_b \to \Lambda_c \ell \bar{\nu}_{\ell})$ and
$R_{L/T}$ can be significantly reduced to $\simeq 12 \%$ and $\simeq 1 \%$,
respectively, by integrating the differential decay rates up to $\omega
\simeq 1.2$. This could allow in particular to extract the $CKM$ matrix
elements $|V_{bc}|$ with a theoretical uncertainty of $\simeq 6 \%$, which
is comparable with present uncertainties obtained from exclusive
semileptonic $B$-meson decays \cite{bigi}.

\indent We want to point out that our estimates of the theoretical errors
include the uncertainties arising both from the model dependence of the $IW$
form factor and the first-order power corrections. In the case of the
longitudinal asymmetry and the (partially integrated) longitudinal to
transverse decay ratio our uncertainties turn out to be remarkably small.
Therefore, provided the effects of higher-order power corrections are small,
the experimental determination of these two quantities is a very interesting tool for testing the $SM$ and for investigating possible New Physics.

\vspace{0.5cm}

\section{Light-Front Calculation of the Isgur-Wise Form Factor}

\indent In this section we just briefly remind the main points of the
light-front procedure for the calculation of the $\Lambda_Q$-type $IW$ form factor $\xi(\omega)$ (see Ref. \cite{plb98} for more general cases) .

\indent The spin structure of $\Lambda_Q$ baryons can be represented by an
antisymmetric (with respect to the two light quarks $q$ and $q'$) total spin-$\frac{1}{2}$ state \cite{iw}. In the so-called "velocity basis" the matrix elements of the vector current $J^{\mu} = \bar{Q} \gamma^{\mu} Q$ between $\Lambda_Q$ states have the following form
 \be
    {\cal{V}}_{s' s}^{\mu} & \equiv & \langle \Psi^{1/2 s'} | \bar{Q}
    \gamma^{\mu} Q | \Psi^{1/2 s} \rangle = F_1(\omega) ~ \bar{u}(P', s')
    \gamma^{\mu} u(P, s) + \nonumber \\ & + & \left [ F_2(\omega) ~ v^{\mu}
    + F_3(\omega) ~ {v'}^{\mu} \right ] ~ \bar{u}(P', s') u(P, s)
    \label{eq:vector}
 \ee
where $u(P, s)$ is a Dirac spinor (normalised as $\bar{u} u = 1$), $P =
M_{\Lambda} v$ and $P' = P + q = M_{\Lambda} v'$, with $M_{\Lambda}$ being
the $\Lambda$-type baryon mass and $q$ the four-momentum transfer. Since in
Eq. (\ref{eq:vector}) we are considering an elastic process, the
four-momentum transfer squared is given by $q^2 = 2 M_{\Lambda}^2 \cdot (1 -
\omega)$ and has space-like values ($q^2 \leq 0$). In the heavy-quark limit
($m_Q \to \infty$) the following $HQS$ relations hold \cite{iw}
 \be
    \mbox{lim}_{m_Q \to \infty} F_1(\omega) & = &  \xi(\omega) \nonumber \\
    \mbox{lim}_{m_Q \to \infty} F_2(\omega) & = & \mbox{lim}_{m_Q \to
    \infty} F_3(\omega) = 0
    \label{eq:HQL}
 \ee
where the $IW$ form factor $\xi(\omega)$ must satisfy the model-independent normalisation $\xi(1) = 1$  at the zero-recoil point $\omega = 1$.

\indent In the light-front formalism all the hadronic form factors
corresponding to a conserved current can always be expressed in terms of the
matrix elements of the {\em plus} component of the current, $J^+ \equiv J^0
+ \hat {n} \cdot \vec J$ (where $\hat{n}$ is the spin-quantization axis); moreover, a reference frame where $q^+ = 0$ is adopted, which allows to suppress the contribution arising from the so-called $Z$-graph (pair creation from the vacuum) at any value of the heavy-quark mass. Thus, from Eq. (\ref{eq:vector}) one easily gets
 \be
    F_1(\omega) & = & {1 \over 2} \mbox{Tr}\left\{ {{\cal{V}}^+ \over 2P^+}
    \right\} + {M_{\Lambda} \over \sqrt{-q^2}} \mbox{Tr}\left\{ {{\cal{V}}^+ 
    i \sigma_2 \over 2P^+} \right\}
    \label{eq:F1} \\
    F_2(\omega) & = & F_3(\omega) = - {M_{\Lambda} \over 2 \sqrt{-q^2}} 
    \mbox{Tr} \left\{ {{\cal{V}}^+ i \sigma_2 \over 2P^+} \right\}
    \label{eq:F2F3}
 \ee
where $\tilde{P} \equiv (P^+, \vec{P}_{\perp}) = \tilde{p}_Q + \tilde{p}_q +
\tilde{p}_{q'}$ is the $LF$ baryon momentum (with $P^+ = \sqrt{M_{\Lambda}^2
- q^2 / 4}$) and $\tilde{p}_i$ the quark one. Finally, the subscript $\perp$
indicates the projection perpendicular to the spin quantization axis.

\indent Disregarding for simplicity the colour and flavour degrees of freedom and limiting ourselves to $S$-wave baryons, the light-front $\Lambda_Q$ wave function can be written as
 \be
    \langle \{ \xi_i \vec{k}_{i \perp}; \sigma_i \}|\Psi^{1/2 s} \rangle = 
    \sqrt{{E_Q E_q E_{q'} \over M_0 \xi_Q \xi_q \xi_{q'}}} \sum_{\{\sigma'_i 
    \}} \langle \{ \sigma_i \} | {\cal{R}}^{\dag}(\{ \vec{k}_i; m_i \}) | \{ 
    \sigma'_i \} \rangle ~ \rangle \Phi^{1/2 s}(\{\sigma'_i\}) ~ 
    w_{(Qqq')}(\vec p, \vec k)
    \label{eq:LF_wf}
 \ee
where $\xi_i = p_i^+ / P^+$ and $\vec{k}_{i \perp} = \vec{p}_{i \perp} -
\xi_i \vec{P}_{\perp}$ are the intrinsic $LF$ variables, $M_0^2 = \sum_{i =
Q, q, q'} (k_{i \perp}^2 + m_i^2) / \xi_i$, $E_i = \sqrt{m_i^2 +
|\vec{k}_i|^2 }$, with  $\vec{k}_i \equiv ( \vec{k}_{i \perp}, k_{in})$ and
$k_{in} = {1 \over 2} ~ \left ( \xi_i M_0 - { {k_{i \perp}^2 + m_i^2} \over
\xi_i M_0} \right )$. Moreover, in Eq. (\ref {eq:LF_wf}) the curly braces
$\{ ~~ \}$ mean a list of indices corresponding to $i = Q, q, q'$;
$\sigma'_i$ indicates the third component of the quark spin;
${\cal{R}}(\{\vec{k}_i; m_i \}) \equiv \prod_{j = Q, q, q'} R_M(\vec{k}_j,
m_j)$ is the product of the individual (generalised) Melosh rotations
$R_M(\vec{k}_j, m_j)$. Finally, $\Phi^{1/2 s}(\{\sigma'_i\}) = \langle 1/2
\sigma'_Q, 00 | 1/2 s \rangle$ $\langle 1/2 \sigma'_q, 1/2 \sigma'_{q'} | 00
\rangle$ is the canonical spin wave function and $w_{(Qqq')}(\vec p, \vec
k)$ is the S-wave radial wave function, where
 \be
    \vec{k} & = & (m_{q'} \vec{k}_q - m_q \vec{k}_{q'}) / (m_q + m_{q'})
    \nonumber \\
    \vec{p} & = & [(m_q + m_{q'}) \vec{k}_Q - m_Q (\vec{k}_q +
    \vec{k}_{q'})] / (m_Q + m_q + m_{q'})
    \label{eq:kp}
 \ee
are the internal momenta conjugate, respectively, to the Jacobian co-ordinates
 \be
    \vec{x} & = & \vec{r}_q - \vec{r}_{q'} \nonumber \\
    \vec{y} & = & \vec{r}_Q - (m_q \vec{r}_q + m_{q'} \vec{r}_{q'}) / (m_q +
    m_{q'})
    \label{eq:xy}
 \ee

\indent In what follows we will consider two functional forms for the radial wave function, namely harmonic oscillator ($h.o.$) and power-law ($p.l.$) ones, viz.
 \be
    \label{eq:ho}
    w_{(Qqq')}^{(h.o.)}(\vec p, \vec k) & = & \left(\frac{1}{\pi \alpha_p}
    \right)^{\frac{3}{4}} e^{-|\vec p|^2 / 2\alpha_p^2} \cdot 
    \left(\frac{1}{\pi \alpha_k}\right)^{\frac{3}{4}} e^{-|\vec k|^2 / 
    2\alpha_k^2} \\ 
    \label{eq:pl} 
    w_{(Qqq')}^{(p.l.)}(\vec p, \vec k) & =  & \frac{N_p}{(\alpha_p^2 + 
    p^2)^{n_p}} \cdot \frac{N_k}{(\alpha_k^2 + k^2)^{n_k}}
 \ee
where $N_{k(p)} \equiv \sqrt{4 \Gamma[2 n_{k(p)}] / \sqrt{\pi} \alpha_{k(p)}^{3 - 4 n_{k(p)}} \Gamma[2 n_{k(p)} - \frac{3}{2}]}$ and $\Gamma(n)$ is the Euler $\Gamma$-function.

\indent We have evaluated the right-hand side of Eqs.
(\ref{eq:F1}-\ref{eq:F2F3}) using the three-quark wave function given by Eq.
(\ref{eq:LF_wf}) and adopting the $h.o.$ (\ref {eq:ho}) and $p.l.$ (\ref
{eq:pl}) radial functions. The numerical integrations, involving
six-dimensional integrals, have been performed through a well-established
Monte Carlo procedure \cite{VEGAS}. The heavy-quark limit ($m_Q \to \infty$)
has been obtained by increasing the value of the heavy-quark mass until full
convergence of the calculated form factors is reached. We have found (see
Ref. \cite{plb98}) that the $HQS$ relations (\ref{eq:HQL}) are fulfilled at
any value of $\omega$. Thus, the $IW$ form factors $\xi(\omega)$, is given by
 \be
    \xi(\omega)  =  {1 \over 2} \mbox{lim}_{m_Q \to \infty} \mbox{Tr}\left\{ 
    {{\cal{V}}^+ \over 2P^+} \right\}
    \label{eq:csi0}
 \ee

\vspace{0.5cm}

\section{The Isgur-Wise Form Factor and the Baryon Structure}

\indent The behaviour of the calculated $IW$ form factor $\xi(\omega)$
depends on the two parameters $\alpha_k$ and $\alpha_p$, appearing in the
radial wave functions (\ref{eq:ho}-\ref{eq:pl}). To investigate such a dependence, let us consider the non-relativistic baryon size $r_B$, defined as 
 \be
    r_B & \equiv & \sqrt{<r^2>_B} \equiv \sqrt{\sum_{i=Q,q,q'} < | \vec r_i 
    - \vec R_{c.m.} |^2 >} \nonumber \\
    & & \to_{m_Q \to \infty} \sqrt{\frac{1 + \eta^2}{(1 + \eta)^2} <x^2> + 2 
    <y^2>}
    \label{eq:size}
 \ee
where $\eta \equiv m_q / m_{q'}$ and $\vec R_{c.m.} = \sum_{i=Q,q,q'} m_i \vec r_i / \sum_{i=Q,q,q'} m_i$. Since $<x^2>$ and $<y^2>$ are proportional to $1 / \alpha_k^2$ and $1 / \alpha_p^2$, respectively, the baryon size $r_B$ can be easily written as a suitable combination of the parameters $\alpha_p$ and $\alpha_k$. For instance, in the case of the $h.o.$ wave function one immediately gets $r_B = \sqrt{\frac{3}{\alpha_p^2} + \frac{1 + \eta^2}{(1 + \eta)^2} \frac{3}{2 \alpha_k^2}}$. 

\indent In the non-relativistic limit the slope of the $IW$ form factor at
the zero-recoil point, $\rho_{IW}^2 \equiv -[d\xi(\omega) / d\omega]_{\omega
= 1}$, is proportional to the square of the baryon size $r_B$, so that when
$\alpha_{k(p)} \to \infty$ one should have $\rho_{IW}^2 \to 0$. However, as
suggested in Ref. \cite{plb98}, the relativistic delocalization of the light
quark positions is expected to increase the slope $\rho_{IW}^2$ and the
departure from the non-relativistic behaviour should appear when the baryon
size $r_B$ becomes much smaller than $\sqrt{\lambda^2_q + \lambda^2_{q'}}$,
where $\lambda_{q(q')} \equiv 1/ m_{q(q')}$ is the constituent-quark Compton
wavelength. For instance, putting $m_q = m_{q'} = 0.22 ~ GeV$, the trigger
value for the delocalization effects should be $r_B < 1.3 ~ fm$.

\indent The above relativistic effects are fully reflected in our
light-front calculations for both the $h.o.$ \cite{plb98} and $p.l.$ wave
functions; in particular, we have found that the slope $\rho_{IW}^2$ is a
monotonically decreasing function of $\alpha_{p(k)}$ and saturates when
$\alpha_{p(k)}$ become enough large so that $r_B < \sqrt{\lambda^2_q +
\lambda^2_{q'}}$. The interesting point is that the saturation property
holds not only for the slope $\rho_{IW}^2$ (i.e., at small recoils), but
also for the $IW$ form factor itself in the whole $\omega$-range accessible
in the decay $\Lambda_b \to \Lambda_c + \ell \bar{\nu}_{\ell}$ (i.e., $1
\leq \omega \lsim 1.44$), as it is clearly illustrated in Fig. 1. Thus, in
the saturation regime the $IW$ form factor does not depend on the two
parameters $\alpha_k$ and $\alpha_p$ separately, but only on one parameter,
the ratio $\beta$, which we define as
 \be
    \beta \equiv \sqrt{{\langle |\vec p|^2 \rangle \over \langle |\vec k|^2 
    \rangle}} & = & {\alpha_p \over \alpha_k} ~~~~~~~~~~~~~~~~~~~~~~
    \mbox{(h.o.)} 
    \nonumber \\
     & = & {\alpha_p \over \alpha_k} ~ \sqrt{{4n_k - 5 \over 4 n_p - 5}} 
     ~~~~~~~~ \mbox{(p.l.)}
    \label{eq:beta}
 \ee
Since $QCD$ is expected to confine hadrons within distances not larger than
$\sim 1 ~ fm$, we will consider in what follows to be in the saturation
regime, where we stress the $IW$ form factor depends only on the parameter
$\beta$ \footnote{In practice, for any value of $\beta$ we choose the value
of the parameter $\alpha_k$ (or equivalently $\alpha_p$) enough large so
that $r_B << \sqrt{\lambda^2_q + \lambda^2_{q'}}$.}. The physical meaning of
this parameter has been already discussed in Ref. \cite{plb98}. In
co-ordinate space one gets $\beta^2 \sim ~ <x^2> / <y^2>$ and, therefore,
the two limiting cases, $\beta << 1$ and $\beta >> 1$, correspond to a
diquark-like and collinear-type configurations, respectively (see Fig. 2 of
Ref. \cite {plb98}). There is however still another parameter, namely the
ratio of the light-quark masses $\eta = m_q / m_{q'}$. The sensitivity of
the calculated $IW$ function to the value of $\eta$ is reported in Fig. 2.
It can clearly be seen that in the light $u, d, s$ spectator sector (where
basically $0.5 \lsim \eta \leq 1$) there is no significant dependence of
$\xi(\omega)$ on the particular value of $\eta$. This is not surprising,
because in the saturation regime the $IW$ form factor is dominated by
relativistic effects, which means that the typical spectator-quark momenta
are large compared with the spectator-quark masses and therefore the
specific values of the latter play a minor role.

\indent Let us now address the issue of the sensitivity of the calculated
$IW$ form factor to the choice of the $Q q q'$ radial wave function. In Fig.
3 we show the $IW$ function calculated at various values of $\beta$ using
the wave functions (\ref{eq:ho}) and (\ref{eq:pl}). It turns out that for
each value of $\beta$ the $IW$ form factors corresponding to different
radial wave functions are packed together in a narrow band, while the change
in $\beta$ affects more heavily the behaviour of the $IW$ form factor.
Therefore we can state that within the $\omega$-range accessible in the
$\Lambda_b \to \Lambda_c + \ell \bar{\nu}_{\ell}$ decay ($1 \leq \omega
\lsim 1.44$) the $\omega$-dependence of the $IW$ function $\xi(\omega)$ is mainly governed by the value of only one parameter, $\beta$, and almost independent of the choice of the particular functional form of the radial wave function.

\indent The above feature is relevant because any constraint on the value of
$\beta$ leads immediately to a constraint on the shape of the $IW$ function
$\xi(\omega)$ in the {\em full} $\omega$-range. In other words, if our
calculated $IW$ function is expanded in a series of powers of ($\omega -
1$), the coefficients of this expansion are not independent each other, but
their values are related to the value of $\beta$. In order to constrain
$\beta$ our predictions for $\xi(\omega)$ corresponding to various values of
$\beta$ are compared in Fig. 4(a) with the recent lattice $QCD$ results of
Ref. \cite{UKQCD}, which have been obtained only at low values of the
recoil. It can be seen that: ~ i) there is a sharp sensitivity to the value
of $\beta$, so that our predictions for $\xi(\omega)$ cover a quite large
range of values; however, an upper bound on $\xi(\omega)$, corresponding to
$\beta \to \infty$ and represented by the solid line\footnote{We have
checked that the value $\beta = 100$ is fully representative of the limiting
case $\beta \to \infty$ for the calculation of the $IW$ form factor
$\xi(\omega)$.}, emerges in our $LF$ quark model; ~ ii) only the
calculations with $\beta \gsim 2$ can reproduce all the lattice points
within the quoted errors. It should be noted that the above limits on
$\beta$ yield an allowed range for the $IW$ function which is narrower than
the spread of the lattice points themselves.

\indent The same range of values for $\beta$ is also suggested by the
comparison with the results of Ref. \cite{gupta}, where dispersive bounds
for the slope $\rho_{IW}^2$ and the curvature $2c_{IW} \equiv
[d^2\xi(\omega) / d\omega^2]_{\omega = 1}$ of the $\Lambda_Q$-type $IW$ form
factor at the zero-recoil point have been derived. It should be mentioned
however that the reliability of the dispersive bounds may be plagued by the
effects of the so-called anomalous thresholds (see, e.g., Ref. \cite{thr}).
With this {\em caveat} in mind, the theoretically allowed domain in the
$\rho_{IW}^2 - c_{IW}$ plane \cite{gupta}, is shown in Fig. 4(b) together
with our results corresponding to various values of $\beta$. The dispersive
bounds suggest a quite strong correlation among $\rho_{IW}^2$ and $c_{IW}$,
which is indeed reproduced in our calculations only for $\beta \gsim 2$.
In conclusion, the results presented in Fig. 4 imply the dominance of
collinear-type configurations in the structure of $Qqq'$ baryons.

\indent In what follows we will estimate the model dependence due to the
non-precise knowledge of the $Q q q'$ wave function using the results of the
calculations of the $IW$ form factor (\ref{eq:csi0}), obtained adopting the
$h.o.$ radial function (\ref{eq:ho}) at $\beta = 2$ and the $p.l.$ function
(\ref{eq:pl}) with $n_p = n_k = 2$ at $\beta = 100$. These two form factors,
which will be denoted by $\xi_L(\omega)$ and $\xi_U(\omega)$, represent our
lower and upper bounds to the $IW$ form factor $\xi(\omega)$, i.e.
$\xi_L(\omega) \leq \xi(\omega) \leq \xi_U(\omega)$, corresponding to
constrain the shape of $\xi(\omega)$ by the dashed and solid lines in Fig.
4(a). A simple polynomial fit for the $\omega$-dependence of $\xi_L(\omega)$
and $\xi_U(\omega)$ can be found in the Appendix. From Eqs.
(\ref{eq:quartic}-\ref{eq:parms}) of the Appendix the slope of the $IW$
function at the zero-recoil point results to be $\rho_{IW}^2 = 1.35 \pm
0.55$.

\vspace{0.5cm}

\section{Analysis of the $\Lambda_b \to \Lambda_c + \ell \bar{\nu}_{\ell}$ decay}

\indent The constraints on the shape of $\xi(\omega)$ obtained in the
previous section can be used to reduce the model-dependence uncertainty in the evaluation of the (partially integrated) exclusive semileptonic branching ratio, defined as 
 \be
    Br_{\Lambda_b \to \Lambda_c \ell \bar{\nu}_{\ell}}(\omega_{max}) \equiv
    \tau_{\Lambda_b} \int_1^{\omega_{max}} d\omega ~ {d\Gamma \over 
    d\omega}(\Lambda_b \to \Lambda_c \ell \bar{\nu}_{\ell})
    \label{eq:Br}
 \ee
where $\tau_{\Lambda_b}$ is the $\Lambda_b$ mean lifetime ($\tau_{\Lambda_b}
= 1.24 \pm 0.08 ~ ps$ \cite{PDG98}) and $d\Gamma / d\omega$ is the
differential decay rate for the $\Lambda_b \to \Lambda_c + \ell
\bar{\nu}_{\ell}$ process. As for the latter, one has (see, e.g., Ref. \cite{korner})
 \be
    {d\Gamma \over d\omega}(\Lambda_b \to \Lambda_c \ell \bar{\nu}_{\ell}) 
    = {d\Gamma_L \over d\omega}(\Lambda_b \to \Lambda_c \ell
    \bar{\nu}_{\ell}) ~ + ~ {d\Gamma_T \over d\omega}(\Lambda_b \to
    \Lambda_c \ell \bar{\nu}_{\ell})
    \label{eq:Gamma}
 \ee
with the longitudinal ($L$) and transverse ($T$) parts given by
 \be 
    \label{eq:GammaL}
    {d\Gamma_L \over d\omega}(\Lambda_b \to \Lambda_c \ell 
    \bar{\nu}_{\ell}) & = & \frac{G_F^2}{(2\pi)^3} |V_{bc}|^2  \frac{q^2 
    p_{\Lambda_c} M_{\Lambda_c}}{12 M_{\Lambda_b}} \left[ |H_{\frac{1}{2}, 
    0}|^2 + |H_{-\frac{1}{2}, 0}|^2  \right] \\
    \label{eq:GammaT}
    {d\Gamma_T \over d\omega}(\Lambda_b \to \Lambda_c \ell 
    \bar{\nu}_{\ell}) & = & \frac{G_F^2}{(2\pi)^3} |V_{bc}|^2 \frac{q^2 
    p_{\Lambda_c} M_{\Lambda_c}}{12 M_{\Lambda_b}} \left[ |H_{\frac{1}{2}, 
    1}|^2 + |H_{-\frac{1}{2}, -1}|^2 \right]
 \ee
where $G_F$ is the Fermi coupling constant, $V_{bc}$ is the relevant $CKM$ matrix element and $p_{\Lambda_c} = M_{\Lambda_c} \sqrt{\omega^2 - 1}$ is the momentum of the daughter baryon $\Lambda_c$ in the $\Lambda_b$ rest frame. Within the $SM$ the helicity amplitudes are given by 
 \be 
    \label{eq:h.a.1}
    H_{\lambda_c, \lambda_W} = H_{\lambda_c, \lambda_W}^V - H_{\lambda_c, 
    \lambda_W}^A 
 \ee
where $H^V$ and $H^A$ are the vector ($V$) and axial-vector ($A$) helicity amplitudes, respectively, and $\lambda_c$ and $\lambda_W$ indicate the helicity of the daughter baryon and the one of the $W$-boson, respectively. The vector and axial- vector helicity amplitudes can be expressed in terms of the vector and axial-vector form factors as \cite{korner}
 \be 
    H_{\frac{1}{2},1}^{V,A} & = & -2 \sqrt{M_{\Lambda_b} M_{\Lambda_c} 
    (\omega \mp 1)} F_1^{V,A}(\omega) \nonumber \\
    H_{\frac{1}{2},0}^{V,A} & = & \frac{1}{\sqrt{q^2}} \sqrt{2 M_{\Lambda_b}
    M_{\Lambda_c} (\omega \mp 1) } \left[ (M_{\Lambda_b} \pm M_{\Lambda_c})
    F_1^{V,A} (\omega) \pm M_{\Lambda_c}(\omega \pm 1) F_2^{V,A} (\omega) 
    \right. \nonumber \\
    & & \left. \pm M_{\Lambda_b}(\omega \pm 1) F_3^{V,A} (\omega) \right]
    \label{eq:h.a.2}
 \ee
where the upper and lower signs stand for the vector ($V$) and the
axial-vector ($A$) case, respectively, and the amplitudes for negative
values of the helicities can be obtained according to the relation $H_{-\lambda_c, -\lambda_W}^{V(A)} = + ~ (-) ~ H_{\lambda_c, \lambda_W}^{V(A)}$.

\indent In the heavy-quark limit ($m_Q \to \infty$) one has
 \be
    F_1^V(\omega) & = & F_1^A(\omega) = \xi(\omega) \nonumber \\
    F_{2,3}^V(\omega) & = & F_{2,3}^A(\omega) = 0
    \label{eq:HQS}
 \ee
Including first-order $1/m_Q$  corrections to the $HQS$ result (\ref{eq:HQS}) the six form factors $F_i^{V(A)}$ become \cite{korner}
 \be
    F_1^V(\omega) & = & \xi(\omega) + \left({\Lambda \over 2m_c} + 
    {\Lambda \over 2m_b} \right) \left[2 \chi(\omega) + \xi(\omega) \right]
    \nonumber \\
    F_2^V(\omega) & = & - {\Lambda \over m_c} {\xi(\omega) \over 1 + 
    \omega} \nonumber \\
    F_3^V(\omega) & = & - {\Lambda \over m_b} {\xi(\omega) \over 1 +
    \omega} \nonumber \\
    F_1^A(\omega) & = & \xi(\omega) + \left({\Lambda \over 2m_c} + 
    {\Lambda \over 2m_b} \right) \left[2 \chi(\omega) + \xi(\omega) {\omega
    - 1 \over \omega + 1} \right] \nonumber \\
    F_2^A(\omega) & = & - {\Lambda \over m_c} {\xi(\omega) \over 1 +
    \omega} \nonumber \\
    F_3^A(\omega) & = & {\Lambda \over m_b} {\xi(\omega) \over 1 +
    \omega}
    \label{eq:first-o} 
    \ee 
where $\Lambda \equiv M_{\Lambda_Q} - m_Q + O(1 / m_Q)$ is the binding energy of the baryon in the heavy-quark limit and $\chi(\omega)$ is an (unknown) sub-leading function subject to the condition $\chi(1) = 0$.

\indent The effect of radiative corrections on the $HQS$ relations (\ref{eq:HQS}) is to relate the vector and axial-vector form factors to the (renormalized) $IW$ function $\xi(\omega)$ as follows (see, e.g., Ref. \cite{neubert})
 \be
    F_i^V(\omega) = C_i^V(\omega) ~ \xi(\omega) \nonumber \\
    F_i^A(\omega) = C_i^A(\omega) ~ \xi(\omega)
    \label{eq:HQS+pQCD}
 \ee
where $C_i^{V(A)}(\omega)$ with $i = 1, 2, 3$ are renormalization-group invariant coefficients, which have been calculated and tabulated in Ref. \cite{neubert}. A convenient parametrized form of the $\omega$-dependence of all the six short-distance coefficients $C_i^{V(A)}(\omega)$ can be found in the Appendix. The inclusion of both radiative and $1 / m_Q$ corrections leads to the following expressions \cite{neubert}
 \be
    F_1^V(\omega) & = & C_1^V(\bar{\omega}) \left\{ \xi(\omega) + 
    \left( {\Lambda \over 2m_c} + {\Lambda \over 2m_b} \right) \left[2 
    \chi(\omega) + \xi(\omega) \right] \right\} \nonumber \\
    F_2^V(\omega) & = & C_2^V(\bar{\omega}) \xi(\omega) + 
    {\Lambda \over 2m_b} C_2^V(\bar{\omega}) \left[ 2 \chi(\omega) +
    \xi(\omega) {3 \omega - 1 \over 1 + \omega} \right] + \nonumber \\
    & & {\Lambda \over 2m_c} \left\{ C_2^V(\bar{\omega}) \left[2 
    \chi(\omega) + \xi(\omega) {\omega - 1 \over 1 + \omega} \right] -
    {2 \over 1 + \omega} \left[C_1^V(\bar{\omega}) + C_3^V(\bar{\omega}) 
    \right] \xi(\omega) \right\} \nonumber \\
    F_3^V(\omega) & = & C_3^V(\bar{\omega}) \xi(\omega) + 
    {\Lambda \over 2m_c} C_3^V(\bar{\omega}) \left[ 2 \chi(\omega) +
    \xi(\omega) {3 \omega - 1 \over 1 + \omega} \right] + \nonumber \\
    & & {\Lambda \over 2m_b} \left\{ C_3^V(\bar{\omega}) \left[2 
    \chi(\omega) + \xi(\omega) {\omega - 1 \over 1 + \omega} \right] -
    {2 \over 1 + \omega} \left[C_1^V(\bar{\omega}) + C_2^V(\bar{\omega}) 
    \right] \xi(\omega) \right\} \nonumber \\
    F_1^A(\omega) & = & C_1^A(\bar{\omega}) \left\{ \xi(\omega) + 
    \left( {\Lambda \over 2m_c} + {\Lambda \over 2m_b} \right) \left[2 
    \chi(\omega) + \xi(\omega) {\omega - 1 \over 1 + \omega} \right] 
    \right\} \nonumber \\
    F_2^A(\omega) & = & C_2^A(\bar{\omega}) \xi(\omega) + 
    {\Lambda \over 2m_b} C_2^A(\bar{\omega}) \left[ 2 \chi(\omega) +
    \xi(\omega) {3 \omega + 1 \over 1 + \omega} \right] + \nonumber \\
    & & {\Lambda \over 2m_c} \left\{ C_2^A(\bar{\omega}) \left[2 
    \chi(\omega) + \xi(\omega) \right] - {2 \over 1 + \omega} 
    \left[C_1^A(\bar{\omega}) + C_3^A(\bar{\omega}) \right] \xi(\omega) 
    \right\} \nonumber \\
    F_3^A(\omega) & = & C_3^A(\bar{\omega}) \xi(\omega) + 
    {\Lambda \over 2m_c} C_3^A(\bar{\omega}) \left[ 2 \chi(\omega) +
    \xi(\omega) {3 \omega + 1 \over 1 + \omega} \right] + \nonumber \\
    & & {\Lambda \over 2m_b} \left\{ C_3^A(\bar{\omega}) \left[2 
    \chi(\omega) + \xi(\omega) \right] + {2 \over 1 + \omega} 
    \left[C_1^A(\bar{\omega}) - C_2^A(\bar{\omega}) \right] \xi(\omega) 
    \right\}
    \label{eq:first-o+pQCD}
 \ee 
where $\bar{\omega} \equiv \omega + (\Lambda / m_b + \Lambda / m_c) (\omega - 1)$.

\indent In this work we calculate also various partially-integrated asymmetries, namely the longitudinal $a_L(\omega_{max})$ and transverse $a_T(\omega_{max})$ asymmetries, defined as
 \be
     \label{eq:aL}
     a_L(\omega_{max}) & = & {\int_1^{\omega_{max}} d\omega ~ K(\omega) ~ 
     \left[ |H_{\frac{1}{2}, 0}|^2 - |H_{-\frac{1}{2}, 0}|^2  \right] \over 
     \int_1^{\omega_{max}} d\omega ~ K(\omega) ~ \left[ |H_{\frac{1}{2}, 
     0}|^2 + |H_{-\frac{1}{2}, 0}|^2 \right]} \\
     \label{eq:aT}
     a_T(\omega_{max}) & = & {\int_1^{\omega_{max}} d\omega ~ K(\omega) ~ 
     \left[ |H_{\frac{1}{2}, 1}|^2 - |H_{-\frac{1}{2}, -1}|^2 \right] \over 
     \int_1^{\omega_{max}} d\omega ~ K(\omega) ~ \left[ |H_{\frac{1}{2}, 
     1}|^2 + |H_{-\frac{1}{2}, -1}|^2  \right]}
 \ee
where $K(\omega) \equiv \frac{G_F^2}{(2\pi)^3} |V_{bc}|^2 \frac{q^2 p_{\Lambda_c} M_{\Lambda_c}}{12 M_{\Lambda_b}}$, and the ratio of the longitudinal to transverse decay rates $R_{L/T}(\omega_{max})$, viz.
 \be
    R_{L/T}(\omega_{max}) & = & {\int_1^{\omega_{max}} d\omega ~ {d\Gamma_L 
    \over d\omega}(\Lambda_b  \to \Lambda_c \ell \bar{\nu}_{\ell}) \over 
    \int_1^{\omega_{max}} d\omega ~ {d\Gamma_T \over d\omega}(\Lambda_b \to 
    \Lambda_c \ell \bar{\nu}_{\ell})}
    \label{eq:ratioLT}
 \ee
Finally, the (partially integrated) longitudinal $\Lambda_c$ polarisation 
$P_L(\omega_{max})$ (as defined in Ref. \cite{korner}) can be easily obtained in terms of $a_L(\omega_{max})$, $a_T(\omega_{max})$ and  $R_{L/T}(\omega_{max})$ as
 \be
    P_L(\omega_{max}) = {a_T(\omega_{max}) + a_L(\omega_{max}) ~
    R_{L/T}(\omega_{max}) \over 1 + R_{L/T}(\omega_{max})}
    \label{eq:PL}
 \ee
Note that within the $SM$ all the various observables (\ref{eq:aL}-\ref{eq:PL}) are independent of the specific value (and uncertainty) of $|V_{bc}|$.

\indent We have evaluated the exclusive semileptonic branching ratio
(\ref{eq:Br}) and the various observables given by Eqs.
(\ref{eq:aL}-\ref{eq:PL}), using the expressions
(\ref{eq:HQS}-\ref{eq:first-o+pQCD}) for the relevant form factors entering
the helicity amplitudes (\ref{eq:h.a.2}) and adopting the two $IW$ functions
$\xi_L(\omega)$ and $\xi_U(\omega)$, determined in the previous Section. Up
to first-order $1 / m_Q$ corrections, the values of the quark masses $m_b$
and $m_c$ are given by $m_b = M_{\Lambda_b} - \Lambda$ and $m_c =
M_{\Lambda_c} - \Lambda$, with $M_{\Lambda_b} = 5.624 ~ GeV$ and
$M_{\Lambda_c} = 2.285 ~ GeV$ from $PDG$ \cite{PDG98}. Thus, in Eqs.
(\ref{eq:first-o}) and (\ref{eq:first-o+pQCD}) there are only two unknowns,
the subleading function $\chi(\omega)$ and the baryonic binding energy
$\Lambda$. In Ref. \cite{UKQCD} the $m_Q$ dependence of the lattice $QCD$
results has been investigated, obtaining that the value $\Lambda = 0.75_{-13
~ -6}^{+10 ~ +5} ~ GeV$ and $\chi(\omega) \simeq 0$ are consistent with the
lattice points. Moreover, the sub-leading function $\chi(\omega)$ has been
recently calculated using $QCD$ sum rules in Ref. \cite{QCDSR} and within
the quark model of Ref. \cite{IMF}. In both calculations the resulting
$\chi(\omega)$ turns out to be quite small (of the order of a few percent);
in particular it has been found to be negative in \cite{QCDSR}, while both
positive and negative subleading functions have been obtained in \cite{IMF}
according to the values adopted for the quark-model parameters. Thus, we
have calculated the exclusive semileptonic branching ratio and the various
asymmetries using for $\chi(\omega)$ the results labelled $I$ in Ref.
\cite{QCDSR}, but considering either a positive or a negative sign. It turns
out that the branching ratio is modified by about $\pm 2 \%$, while the
effect on the asymmetries is well below $\pm 1 \%$. Therefore, in what
follows we adopt the approximation $\chi(\omega) = 0$ and inflate the
theoretical error of the calculated branching ratio by adding a $\pm 2 \%$
uncertainty.

\indent We have firstly investigated the sensitivity of the observables
(\ref{eq:Br}) and (\ref{eq:aL}-\ref{eq:PL}), integrated in the whole
$\omega$-range accessible in $\Lambda_b \to \Lambda_c + \ell
\bar{\nu}_{\ell}$ (i.e., for $\omega_{max} = \omega_{th} = (M_{\Lambda_b}^2
+ M_{\Lambda_c}^2) / 2 M_{\Lambda_b} M_{\Lambda_c} \simeq 1.44$), to the
radiative corrections in the heavy-quark limit (see Eqs.
(\ref{eq:HQS+pQCD})). Our predictions, corresponding to the average $\pm$
the semi-dispersion of the results obtained using $\xi(\omega) =
\xi_L(\omega)$ and $\xi(\omega) = \xi_U(\omega)$, are reported in Table 1.
It can clearly be seen that $pQCD$ corrections lower $Br(\Lambda_b \to
\Lambda_c \ell \bar{\nu}_{\ell}) \equiv Br_{\Lambda_b \to \Lambda_c \ell
\bar{\nu}_{\ell}}(\omega_{th})$ and $a_T \equiv a_T(\omega_{th})$ by about
$10 \%$, while the other asymmetries $a_L \equiv a_L(\omega_{th})$ and $P_L
\equiv P_L(\omega_{th})$ as well as the ratio $R_{L/T} \equiv
R_{L/T}(\omega_{th})$ are only marginally modified.

\indent Then, since the $HQET$ parameter $\Lambda$ governs the strength of
the $1 / m_Q$ corrections, we have investigated the sensitivity of the
(totally integrated) exclusive semileptonic branching ratio and asymmetries
to the specific value of $\Lambda$ both with and without radiative
corrections (see Eqs. (\ref{eq:first-o+pQCD}) and (\ref{eq:first-o}),
respectively). Our predictions are reported in Tables 2 and 3. It can
clearly be seen that:

\begin{itemize}

\item the first-order $1 / m_Q$ corrections yield an increase of the
calculated $Br(\Lambda_b \to \Lambda_c \ell \bar{\nu}_{\ell})$ (about $10
\%$ at $\Lambda = 0.75 ~ GeV$), which is partially compensated by a
corresponding decrease due to radiative corrections. Thus, radiative plus
first-order power corrections affect only marginally the exclusive
semileptonic branching ratio, so that the model dependence turns out to be
the most important source of uncertainty on $Br(\Lambda_b \to \Lambda_c \ell
\bar{\nu}_{\ell})$;

\item the longitudinal asymmetry $a_L$ is only slightly sensitive to both radiative and $1 / m_Q$ corrections;

\item the effects of radiative and $1 / m_Q$ corrections are opposite in the 
longitudinal to transverse decay ratio $R_{L/T}$, but the total effect is smaller than the uncertainty due to the model dependence;

\item the transverse asymmetry $a_T$ and the longitudinal $\Lambda_c$ polarisation $P_L$ are remarkably affected by radiative and $1/ m_Q$ corrections, while their model dependence is quite smaller.

\end{itemize}

\noindent Similar conclusions hold as well also for the partially integrated
observables at $\omega_{max} < \omega_{th}$, as it can be seen in Table 4 in
the case of the exclusive semileptonic branching ratio $Br_{\Lambda_b \to
\Lambda_c \ell \bar{\nu}_{\ell}}(\omega_{max})$ (see also Fig. 6 later on).
Therefore, taking into account both the model dependence and the sensitivity
to the $1 / m_Q$ corrections in the whole range of values $0 \leq \Lambda
(GeV) \leq 1$, our final predictions (including radiative corrections) are:
$Br(\Lambda_b \to \Lambda_c \ell \bar{\nu}_{\ell}) = (6.3 \pm 1.6) ~ \% ~
|V_{bc} / 0.040|^2 ~ \tau(\Lambda_b) / (1.24 ~ ps)$, $a_L = -0.945 \pm
0.014$, $a_T = -0.62 \pm 0.09$ and $R_{L/T} = 1.57 \pm 0.15$. The
corresponding result for the longitudinal $\Lambda_c$ polarisation is $P_L =
-0.82 \pm 0.05$.

\indent In Fig. 5 the upper (corresponding to $\xi(\omega) = \xi_U(\omega)$)
and lower (corresponding to $\xi(\omega) = \xi_L(\omega)$) results for the
partially integrated exclusive semileptonic branching ratio $Br_{\Lambda_b
\to \Lambda_c \ell \bar{\nu}_{\ell}}(\omega_{max})$, obtained adopting the
value $\Lambda = 0.75 ~ GeV$ in Eq. (\ref{eq:first-o+pQCD}), are shown and
compared with the $HQS$ results (Eqs. (\ref{eq:HQS})) and with the lattice
predictions of Ref. \cite{UKQCD}. It can be seen that our results are always
well within the range of values given by the lattice $QCD$ simulations and
that radiative plus first-order power corrections modify only slightly the
$HQS$ results. If the integration over the recoil is limited to
$\omega_{max} = 1.2$, then the resulting uncertainty on $Br_{\Lambda_b \to
\Lambda_c \ell \bar{\nu}_{\ell}}(\omega_{max})$  reduces significantly to
$\simeq 12 \%$, though at the price of reducing the number of the events by
a factor $\simeq 0.44$ (see Table 2). This implies the possibility to
extract the $CKM$ matrix elements $|V_{bc}|$ with a theoretical uncertainty
of $\simeq 6 \%$, which is comparable with present uncertainties obtained
from exclusive semileptonic $B$-meson decays \cite{bigi}.

\indent The dependence of the various asymmetries (\ref{eq:aL}-\ref{eq:PL})
upon $\omega_{max}$ is illustrated in Table 5 and in Fig. 6, where the
lattice predictions \cite{UKQCD} for $R_{L/T}(\omega_{max})$ are also
reported. All the observables considered do depend on the specific value of
$\omega_{max}$, so that in comparing with (future) data the precise
$\omega$-range of the experiments has to be taken into account. In Fig. 6
the $HQS$ results are also shown for each observable. It turns out that
radiative plus first-order $1 / m_Q$ corrections are relevant for the
transverse asymmetry $a_T(\omega_{max})$ and for the longitudinal
$\Lambda_c$ polarisation $P_L(\omega_{max})$ at any value of $\omega_{max}$,
whereas both $a_L(\omega_{max})$ and $R_{L/T}(\omega_{max})$ are only
marginally affected by radiative plus first-order power corrections.
Moreover, the model dependence on the various asymmetries is generally quite
limited and it reduces as $\omega_{max}$ decreases. In particular, our
uncertainty on $R_{L/T}(\omega_{max})$, which is always much less than the
one presently achievable by lattice $QCD$ calculations, reduces to $\simeq 1
\%$ at $\omega_{max} = 1.2$. To sum up, both the longitudinal asymmetry
$a_L(\omega_{th})$ and the longitudinal to transverse decay ratio
$R_{L/T}(\omega_{max} \simeq 1.2)$ represent very interesting quantities to
be determined experimentally in order to test the $SM$ and to investigate
possible New Physics.

\indent All the conclusions obtained so far are based on Eqs.
(\ref{eq:first-o+pQCD}) for the form factors, i.e. on the inclusion of power
corrections up to first order in $1 / m_Q$. The second-order power
corrections to $F_1^A$ and $F_1^V + F_2^V + F_3^V$ at the zero-recoil point
have been estimated in Ref. \cite{korner_2} and found to be of the order of
a few percent. In the case of the $B \to D^* (D)$ transitions important
effects of second-order power corrections may be present (see, e.g., Ref.
\cite{holdom_1}); however, a (model-dependent) analysis \cite{holdom_2} of
$1 / m_Q^2$ corrections in mesons and baryons suggests that smaller
corrections can be expected in the $\Lambda_b \to \Lambda_c$ transition with
respect to the $B \to D^* (D)$ case, thanks to the absence of hyperfine
splitting effects related to the finite mass of the charm quark.
Nevertheless, we feel it is mandatory to check that our findings
(particularly those for the asymmetries) still survive after the inclusion
of second-order power corrections.

\indent Before closing this Section, we want to present our estimates of the
exclusive semileptonic branching ratio and the asymmetries for the decay
process $\Xi_b \to \Xi_c + \ell \bar{\nu}_{\ell}$, where the light spectator
quarks are a $us$ ($ds$) pair. As already illustrated in Fig. 2, we do not
expect that the $IW$ form factor is modified significantly by the presence
of the $s$ quark mass instead of the $u(d)$ one; this does not imply that
the $IW$ form factors for $\Lambda_Q$- and $\Xi_Q$-type baryons are the
same, because of possible differences in the radial wave functions in the
two heavy systems. In the absence of a more precise knowledge of $SU(3)$
breaking effects, we assume the same range of variation of the parameter
$\beta$, i.e., $\beta \gsim 2$. Moreover, both the binding energy
$\Lambda^{(s)}$ and the physical threshold $\omega_{th}^{(s)} = (M_{\Xi_b}^2
+ M_{\Xi_c}^2) / 2 M_{\Xi_b} M_{\Xi_c}$ have to be determined. Since the $c$
quark mass should be the same in the two decay processes $\Lambda_b \to
\Lambda_c + \ell \bar{\nu}_{\ell}$ and $\Xi_b \to \Xi_c + \ell
\bar{\nu}_{\ell}$, we assume $M_{\Lambda_c} - \Lambda = M_{\Xi_c} - \Lambda
^{(s)}$. Using the experimental value $M_{\Xi_c} = 2.466 ~ GeV$ \cite{PDG98}
we get $\Lambda ^{(s)} = 0.931 ~ GeV$. In an analogous way for the $b$-quark
case, we consider that $M_{\Lambda_b} - \Lambda = M_{\Xi_b} -
\Lambda^{(s)}$; using the previously determined value of $\Lambda^{(s)}$, we
get $M_{\Xi_b} = 5.805 ~ GeV$, which is in reasonable overall agreement with
the lattice $QCD$ results $5.76_{-5 ~ -3}^{+3 ~ +4} ~ GeV$ from Ref.
\cite{UKQCD_M}. Finally, from the obtained values of the $\Xi_Q$-type baryon
masses one has $\omega_{th}^{(s)} \simeq 1.39$. Thus, after including
radiative plus first-order $1 / m_Q$ corrections (\ref{eq:first-o+pQCD}) to
the relevant form factors our estimates are: $Br(\Xi_b \to \Xi_c \ell
\bar{\nu}_{\ell}) = (7.9 \pm 2.0 ) ~ \%$ (at $|V_{bc}| = 0.040$ and
$\tau(\Xi_b) = 1.39 ~ ps$ \cite{PDG98}), $a_L(\Xi_b \to \Xi_c \ell
\bar{\nu}_{\ell}) = -0.947 \pm 0.013$, $a_T(\Xi_b \to \Xi_c \ell
\bar{\nu}_{\ell}) = -0.62 \pm 0.11$, $P_L(\Xi_b \to \Xi_c \ell
\bar{\nu}_{\ell}) = -0.87 \pm 0.05$  and $R_{L/T} (\Xi_b \to \Xi_c \ell
\bar{\nu}_{\ell}) = 1.60 \pm 0.14$.

\vspace{0.5cm}

\section{Conclusions}

\indent The Isgur-Wise form factor relevant for the $\Lambda_b \to \Lambda_c
+ \ell \bar{\nu}_{\ell}$ semileptonic decay has been calculated in the whole
accessible kinematical range adopting a light-front constituent quark model
and using various forms of the $Qqq'$ wave function. It has been shown that
the $IW$ form factor is sensitive to light-quark relativistic delocalization
effects, leading to a saturation property of the form factor as a function
of the (non-relativistic) baryon size (\ref{eq:size}). Moreover, the
behaviour of the $IW$ function turns out to be largely affected by the
baryon structure, being sharply different in the case of diquark or
collinear-type $Qqq'$ configurations. The comparison with recent lattice
$QCD$ calculations \cite{UKQCD} at low recoil suggests the dominance of
collinear-type configurations with respect to diquark-like ones and allows
to put effective constraints on the shape of the $IW$ function in the full
$\omega$-range. Then, the $\Lambda_b \to \Lambda_c + \ell \bar{\nu}_{\ell}$
decay has been investigated including both radiative and first-order $1 /
m_Q$ corrections to the relevant form factors. Our final predictions for the
exclusive semileptonic branching ratio, the longitudinal and transverse
asymmetries, and the longitudinal to transverse decay ratio are:
$Br(\Lambda_b \to \Lambda_c \ell \bar{\nu}_{\ell}) = (6.3 \pm 1.6) ~ \% ~
|V_{bc} / 0.040|^2 ~ \tau(\Lambda_b) / (1.24 ~ ps)$, $a_L = -0.945 \pm
0.014$, $a_T = -0.62 \pm 0.09$ and $R_{L/T} = 1.57 \pm 0.15$, respectively.
It has also been shown that the theoretical uncertainties on $Br(\Lambda_b
\to \Lambda_c \ell \bar{\nu}_{\ell})$ and $R_{L/T}$ can be significantly
reduced to $\simeq 12 \%$ and $\simeq 1 \%$, respectively, by integrating
the differential decay rates up to $\omega \simeq 1.2$. This could allow the
extraction of the $CKM$ matrix elements $|V_{bc}|$ with a theoretical
uncertainty of $\simeq 6 \%$, which is comparable with present uncertainties
obtained from exclusive semileptonic $B$-meson decays. Finally, we stress
that, provided the effects of higher-order power corrections are negligible,
the small uncertainties found for the longitudinal asymmetry and the
(partially integrated) longitudinal to transverse decay ratio make the
experimental determination of these quantities a very interesting tool for
testing the $SM$ and investigating possible New Physics.

\newpage

\section{APPENDIX}

\indent The results of the calculations of the form factors $\xi_L(\omega)$ and $\xi_U(\omega)$ (see Section 3), performed in the whole $\omega$-range accessible in the decay $\Lambda_b \to \Lambda_c + \ell \bar{\nu}_{\ell}$, can be fitted with high accuracy by the following quartic polynomials in the recoil variable $\omega - 1$, viz.
 \be
    \xi_{L(U)}(\omega) = 1 - \hat{\rho}_{L(U)}^2 \cdot (\omega - 1) + 
    \hat{c}_{L(U)} \cdot (\omega - 1)^2 - \hat{d}_{L(U)} \cdot (\omega - 
    1)^3 + \hat{f}_{L(U)} \cdot (\omega - 1)^4
    \label{eq:quartic}
 \ee
where the values of the parameters are:
 \be
    \hat{\rho}_L^2 & = & 1.884, ~~~~~~~~~~ \hat{c}_L = 2.241, ~~~~~~~~~~ 
    \hat{d}_L = 1.626, ~~~~~~~~~~ \hat{f}_L = 0.5143 \nonumber \\
    \hat{\rho}_U^2 & = & 0.8252, ~~~~~~~~ \hat{c}_U = 0.5388, ~~~~~~~~ 
    \hat{d}_U = 0.2594, ~~~~~~~~ \hat{f}_U = 0.06308
    \label{eq:parms}
 \ee

\indent We stress that neither a linear nor a quadratic polynomial is able
to reproduce accurately the $\omega$-behaviour of the form factors
$\xi_L(\omega)$ and $\xi_U(\omega)$ in the whole range of values of the
recoil accessible in the decay $\Lambda_b \to \Lambda_c + \ell
\bar{\nu}_{\ell}$.

\indent The results of the calculations of the short-distance coefficients $C_i^{V(A)}(\omega)$ reported in Ref. \cite{neubert} can be fitted for $\omega \lsim 1.8$ as follows
 \be
    C_1^V(\omega) & = & 1.136 - 0.2978 \cdot (\omega - 1) + 0.01149 \cdot
    (\omega - 1)^2 - 0.03536 \cdot (\omega - 1)^3 \nonumber \\
    C_2^V(\omega) & = & -0.08485 + 0.04645 \cdot (\omega - 1) - 0.02792 
    \cdot (\omega - 1)^2 + 0.01263 \cdot (\omega - 1)^3 \nonumber \\
    C_3^V(\omega) & = & -0.02133 + 0.007972 \cdot (\omega - 1) - 0.001840 
    \cdot (\omega - 1)^2
    \label{eq:CiV}
 \ee
and
 \be
    C_1^A(\omega) & = & 0.9851 - 0.2069 \cdot (\omega - 1)  + 0.04899 \cdot
    (\omega - 1)^2 - 0.001684 \cdot (\omega - 1)^3 \nonumber \\
    C_2^A(\omega) & = & -0.1220 + 0.07378 \cdot (\omega - 1)  - 0.04062 
    \cdot (\omega - 1)^2 + 0.01347 \cdot (\omega - 1)^3 \nonumber \\
    C_3^A(\omega) & = & 0.04203 - 0.02193 \cdot (\omega - 1) + 0.008658 
    \cdot (\omega - 1)^2.
    \label{eq:CiA}
 \ee

\newpage

\vspace{2cm}

\indent {\bf Table 1}. Results for the (totally integrated) exclusive
semileptonic branching ratio $Br(\Lambda_b \to \Lambda_c \ell
\bar{\nu}_{\ell})$ (in $\%$ at $|V_{bc}| = 0.040$ and $\tau(\Lambda_b) =
1.24 ~ ps$ \cite{PDG98}), the longitudinal $a_L$ and transverse $a_T$
asymmetries, the longitudinal daughter-baryon polarisation $P_L$ and the
longitudinal to transverse ratio $R_{L/T}$ obtained for the decay process
$\Lambda_b \to \Lambda_c + \ell \bar{\nu}_{\ell}$ in the $HQS$ limit ($m_Q
\to \infty$) without and with radiative corrections (see Eqs. (\ref{eq:HQS})
and (\ref{eq:HQS+pQCD}), respectively). The values reported are the average
and the semi-dispersion of the results obtained using $\xi(\omega) =
\xi_L(\omega)$ and $ \xi(\omega) = \xi_U(\omega)$ (see text and Appendix).

\vspace{0.5cm}

\begin{center}

{\bf Table 1}

\vspace{0.25cm}

\begin{tabular} {||c ||c ||c ||c ||c ||c ||} \hline
             & $Br(\%)$ & $a_L$ & $a_T$ & $P_L$ & $R_{L/T}$ \\
\hline \hline
$HQS$        & $6.65 \pm 1.41$ & $-0.931 \pm 0.009$ & $-0.488 \pm 0.012$ & $-0.763 \pm 0.018$ & $1.63 \pm 0.12$ \\
\cline{1-6} \cline{1-6}
$HQS + pQCD$ & $6.07 \pm 1.27$ & $-0.939 \pm 0.009$ & $-0.539 \pm 0.012$ & $-0.785 \pm 0.018$ & $1.61 \pm 0.12$ \\ 
\cline{1-6} \cline{1-6}
\hline
\end{tabular}

\end{center}

\vspace{4cm}

\indent {\bf Table 2}. Results for the (totally integrated) exclusive
semileptonic branching ratio $Br(\Lambda_b \to \Lambda_c \ell
\bar{\nu}_{\ell})$ (in $\%$ at $|V_{bc}| = 0.040$ and $\tau(\Lambda_b) =
1.24 ~ ps$ \cite{PDG98}), the longitudinal $a_L$ and transverse $a_T$
asymmetries, the longitudinal daughter-baryon polarisation $P_L$ and the
longitudinal to transverse ratio $R_{L/T}$ obtained at different values of
the binding energy $\Lambda$ without including radiative corrections (see
Eqs. (\ref{eq:first-o})). For the meaning of the errors see Table 1.

\vspace{0.5cm}

\begin{center}

{\bf Table 2}

\vspace{0.25cm}

\begin{tabular} {||c ||c ||c ||c ||c ||c ||} \hline
$\Lambda$ (GeV) & $Br(\%)$ & $a_L$ & $a_T$ & $P_L$ & $R_{L/T}$ \\
\hline \hline
0.00  & $6.65 \pm 1.41$ & $-0.931 \pm 0.009$ & $-0.488 \pm 0.012$ & $-0.763 \pm 0.018$ & $1.63 \pm 0.12$ \\
\cline{1-6} \cline{1-6}
0.25  & $6.81 \pm 1.45$ & $-0.934 \pm 0.009$ & $-0.521 \pm 0.012$ & $-0.777 \pm 0.018$ & $1.63 \pm 0.12$ \\ 
\cline{1-6} \cline{1-6}
0.50  & $7.03 \pm 1.51$ & $-0.937 \pm 0.009$ & $-0.559 \pm 0.013$ & $-0.793 \pm 0.017$ & $1.64 \pm 0.12$ \\ 
\cline{1-6} \cline{1-6}
0.75  & $7.35 \pm 1.58$ & $-0.942 \pm 0.008$ & $-0.604 \pm 0.013$ & $-0.814 \pm 0.016$ & $1.65 \pm 0.12$ \\ 
\cline{1-6} \cline{1-6}
1.00  & $7.82 \pm 1.70$ & $-0.948 \pm 0.008$ & $-0.658 \pm 0.013$ & $-0.839 \pm 0.015$ & $1.67 \pm 0.12$ \\ 
\cline{1-6} \cline{1-6} 
\hline
\end{tabular}

\end{center}

\newpage

\vspace{2cm}

\indent {\bf Table 3}. The same as in Table 2, but including radiative 
corrections (see Eqs. (\ref{eq:first-o+pQCD})). For the meaning of the 
errors see Table 1.

\vspace{0.5cm}

\begin{center}

{\bf Table 3}

\vspace{0.25cm}

\begin{tabular} {||c ||c ||c ||c ||c ||c ||} \hline
$\Lambda$ (GeV) & $Br(\%)$ & $a_L$ & $a_T$ & $P_L$ & $R_{L/T}$ \\
\hline \hline
0.00  & $6.07 \pm 1.27$ & $-0.939 \pm 0.009$ & $-0.539 \pm 0.013$ & $-0.785 \pm 0.017$ & $1.61 \pm 0.12$ \\
\cline{1-6} \cline{1-6}
0.25  & $6.11 \pm 1.27$ & $-0.941 \pm 0.008$ & $-0.571 \pm 0.013$ & $-0.798 \pm 0.017$ & $1.59 \pm 0.12$ \\ 
\cline{1-6} \cline{1-6}
0.50  & $6.18 \pm 1.29$ & $-0.943 \pm 0.008$ & $-0.604 \pm 0.013$ & $-0.812 \pm 0.016$ & $1.57 \pm 0.11$ \\ 
\cline{1-6} \cline{1-6}
0.75  & $6.28 \pm 1.31$ & $-0.946 \pm 0.008$ & $-0.651 \pm 0.013$ & $-0.830 \pm 0.015$ & $1.55 \pm 0.11$ \\ 
\cline{1-6} \cline{1-6}
1.00  & $6.44 \pm 1.34$ & $-0.950 \pm 0.007$ & $-0.701 \pm 0.013$ & $-0.851 \pm 0.013$ & $1.53 \pm 0.11$ \\ 
\cline{1-6} \cline{1-6} 
\hline
\end{tabular}

\end{center}

\vspace{4cm}

\indent {\bf Table 4}. Values of the (partially integrated) exclusive
semileptonic branching ratio $Br_{\Lambda_b \to \Lambda_c \ell
\bar{\nu}_{\ell}}(\omega_{max})$ in $\%$ versus $\omega_{max}$ (see Eq.
(\ref{eq:Br})), calculated at $|V_{bc}| = 0.040$ and $\tau(\Lambda_b) = 1.24
~ ps$ \cite{PDG98} for the decay process $\Lambda_b \to \Lambda_c + \ell
\bar{\nu}_{\ell}$. The columns labelled $HQS$, $HQS + pQCD$, $HQS + 1 / m_Q$
and $HQS + pQCD + 1 / m_Q$ correspond to the heavy-quark limit $m_Q \to
\infty$ (Eqs. (\ref{eq:HQS})), to the inclusion of radiative corrections
(Eqs. (\ref{eq:HQS+pQCD})) and of the first-order $1 / m_Q$ corrections, as
given by Eqs. (\ref{eq:first-o}) and (\ref{eq:first-o+pQCD}) with $\Lambda =
0.75 ~ GeV$, respectively. For the meaning of the errors see Table 1.

\vspace{0.5cm}

\begin{center}

{\bf Table 4}

\vspace{0.25cm}

\begin{tabular} {||c ||c ||c ||c ||c ||} \hline
$\omega_{max}$ & $HQS$ & $HQS + pQCD$ & $HQS + 1 / m_Q$ & $HQS + pQCD + 1 / m_Q$ \\
\hline \hline
1.05  & $0.43 \pm 0.01$ & $0.42 \pm 0.01$ & $0.44 \pm 0.01$ & $0.43 \pm 0.01$ \\
\cline{1-5} \cline{1-5}
1.10  & $1.15 \pm 0.07$ & $1.10 \pm 0.07$ & $1.19 \pm 0.07$ & $1.13 \pm 0.07$ \\ 
\cline{1-5} \cline{1-5}
1.15  & $1.97 \pm 0.17$ & $1.88 \pm 0.17$ & $2.08 \pm 0.19$ & $1.94 \pm 0.17$ \\ 
\cline{1-5} \cline{1-5}
1.20  & $2.84 \pm 0.33$ & $2.69 \pm 0.31$ & $3.03 \pm 0.35$ & $2.79 \pm 0.32$ \\ 
\cline{1-5} \cline{1-5}
1.25  & $3.72 \pm 0.52$ & $3.49 \pm 0.48$ & $4.01 \pm 0.56$ & $3.64 \pm 0.50$ \\ 
\cline{1-5} \cline{1-5} 
1.30  & $4.57 \pm 0.74$ & $4.26 \pm 0.68$ & $4.97 \pm 0.81$ & $4.44 \pm 0.71$ \\ 
\cline{1-5} \cline{1-5}
1.35  & $5.39 \pm 0.98$ & $4.99 \pm 0.89$ & $5.90 \pm 1.09$ & $5.19 \pm 0.93$ \\ 
\cline{1-5} \cline{1-5}
1.40  & $6.16 \pm 1.24$ & $5.65 \pm 1.12$ & $6.79 \pm 1.38$ & $5.87 \pm 1.16$ \\ 
\cline{1-5} \cline{1-5}
1.44  & $6.65 \pm 1.41$ & $6.07 \pm 1.27$ & $7.35 \pm 1.58$ & $6.28 \pm 1.31$ \\ 
\cline{1-5} \cline{1-5}
\hline
\end{tabular}

\end{center}

\newpage

\vspace{2cm}

\indent {\bf Table 5}. Values of the (partially integrated) asymmetries
$a_L(\omega_{max})$ (Eq. (\ref{eq:aL})) and $a_T(\omega_{max})$ (Eq.
(\ref{eq:aT})), the longitudinal daughter-baryon polarisation
$P_L(\omega_{max})$ (Eq. (\ref{eq:PL})) and the (partially integrated) ratio
of longitudinal to transverse decay rate $R_{L/T}(\omega_{max})$ (Eq.
(\ref{eq:ratioLT})) versus $\omega_{max}$ for the decay process $\Lambda_b
\to \Lambda_c + \ell \bar{\nu}_{\ell}$. Both the radiative and the first-order $1 / m_Q$ corrections, given by Eqs. (\ref{eq:first-o+pQCD}) with $\Lambda = 0.75 ~ GeV$, are included. For the meaning of the errors see Table 1.

\vspace{0.5cm}

\begin{center}

{\bf Table 5}

\vspace{0.25cm}

\begin{tabular} {||c ||c ||c ||c ||c ||} \hline
$\omega_{max}$ & $a_L$ & $a_T$ & $P_L$ & $R_{L/T}$ \\
\hline \hline
1.01  & $-0.282 \pm 0.001$ & $-0.159 \pm 0.001$ & $-0.201 \pm 0.001$ & $0.515 \pm 0.001$ \\
\cline{1-5} \cline{1-5}
1.05  & $-0.571 \pm 0.002$ & $-0.338 \pm 0.001$ & $-0.423 \pm 0.002$ & $0.576 \pm 0.001$ \\
\cline{1-5} \cline{1-5}
1.10  & $-0.725 \pm 0.004$ & $-0.451 \pm 0.003$ & $-0.557 \pm 0.006$ & $0.657 \pm 0.002$ \\ 
\cline{1-5} \cline{1-5}
1.15  & $-0.808 \pm 0.005$ & $-0.521 \pm 0.005$ & $-0.643 \pm 0.006$ & $0.745 \pm 0.005$ \\ 
\cline{1-5} \cline{1-5}
1.20  & $-0.858 \pm 0.006$ & $-0.569 \pm 0.007$ & $-0.701 \pm 0.007$ & $0.842 \pm 0.011$ \\ 
\cline{1-5} \cline{1-5}
1.25  & $-0.891 \pm 0.007$ & $-0.603 \pm 0.009$ & $-0.743 \pm 0.009$ & $0.951 \pm 0.019$ \\ 
\cline{1-5} \cline{1-5} 
1.30  & $-0.913 \pm 0.007$ & $-0.626 \pm 0.010$ & $-0.775 \pm 0.011$ & $1.088 \pm 0.032$ \\ 
\cline{1-5} \cline{1-5}
1.35  & $-0.929 \pm 0.008$ & $-0.641 \pm 0.012$ & $-0.800 \pm 0.012$ & $1.226 \pm 0.051$ \\ 
\cline{1-5} \cline{1-5}
1.40  & $-0.940 \pm 0.008$ & $-0.649 \pm 0.013$ & $-0.819 \pm 0.014$ & $1.404 \pm 0.081$ \\ 
\cline{1-5} \cline{1-5}
1.44  & $-0.946 \pm 0.008$ & $-0.651 \pm 0.013$ & $-0.830 \pm 0.015$ & $1.548 \pm 0.110$ \\ 
\cline{1-5} \cline{1-5}
\hline
\end{tabular}

\end{center}

\newpage

\begin{figure}[htb]

\centerline{\epsfxsize=12cm \epsfig{file=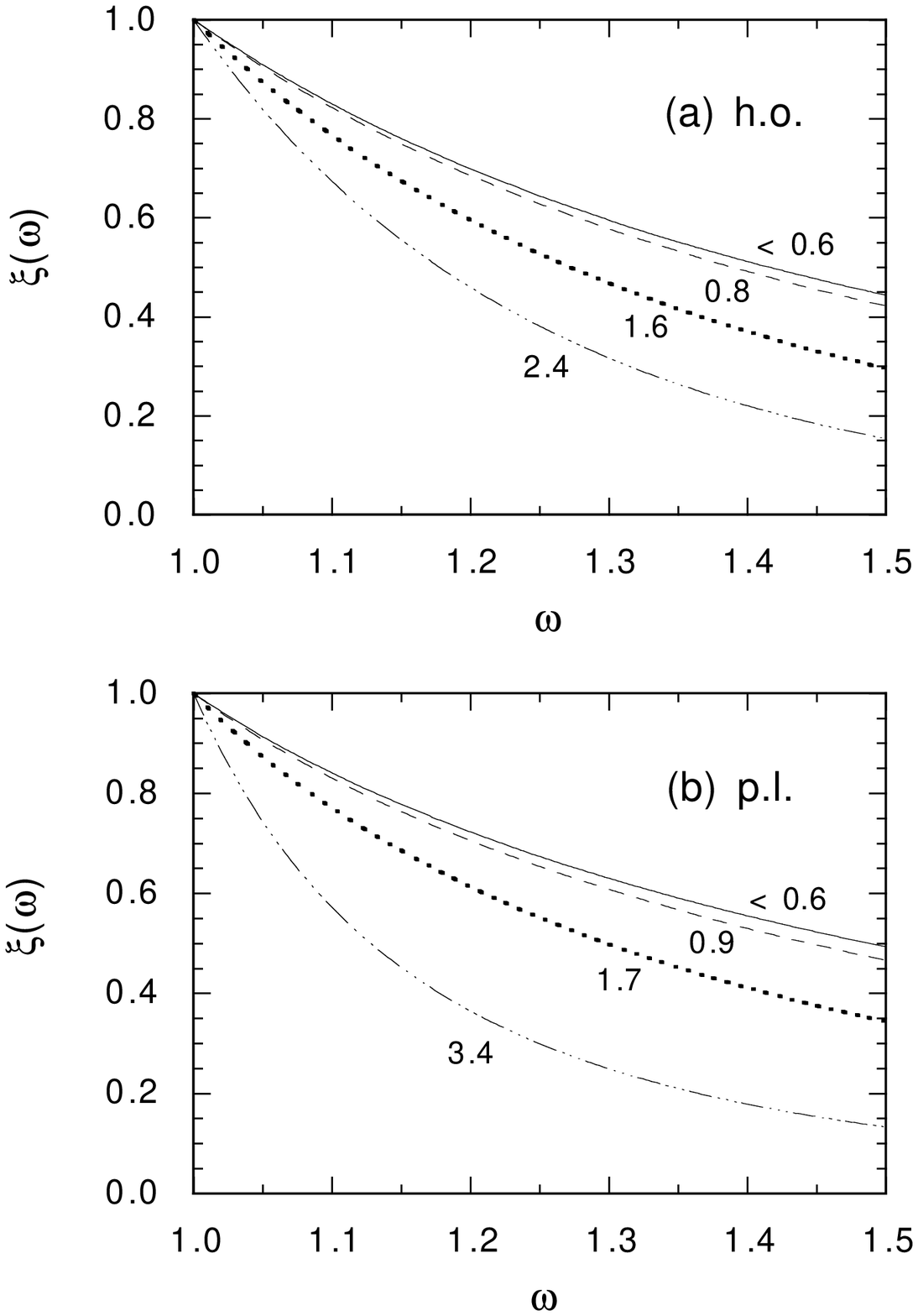}}

\end{figure}

\vspace{0.5cm}

\noindent {\bf Figure 1}. The $IW$ form factor $\xi(\omega)$ calculated 
using $h.o.$ (a) and $p.l.$ (b) wave functions versus $\omega$ (in (b) the 
value $n_p = n_k = 2$ has been considered). The various lines correspond to 
different values of the parameter $\alpha_k$, while the ratio $\beta$ (Eq.
(\ref{eq:beta})) is kept fixed at $\beta = 2$. For each curve the value of
the baryon size $r_B$ (Eq. (\ref{eq:size})) in fm is reported. Finally, for
the spectator-quark masses the value $m_q = m_{q'} = 0.22 ~ GeV$ is adopted.

\newpage

\begin{figure}[htb]

\centerline{\epsfxsize=16cm \epsfig{file=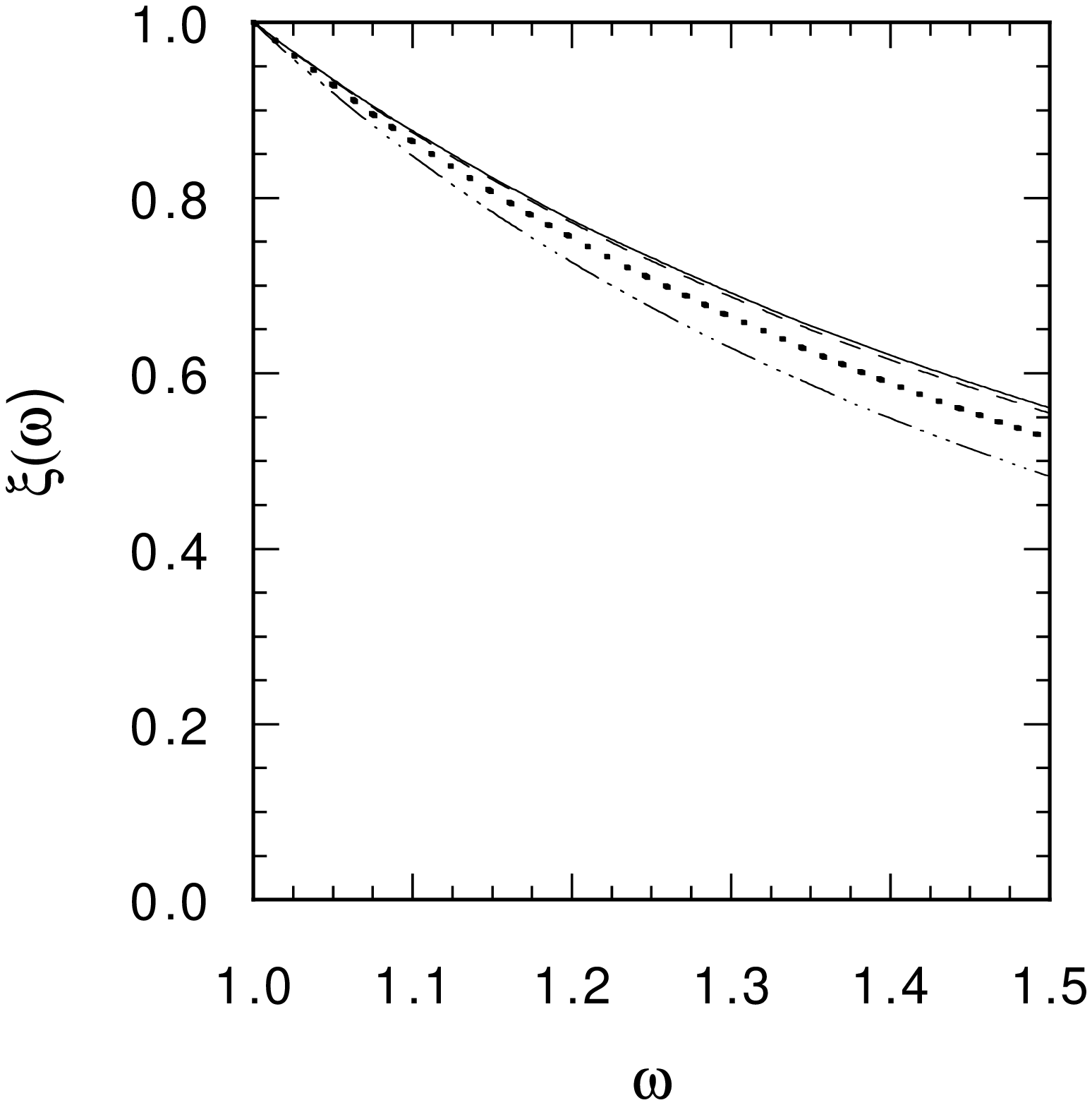}}

\end{figure}

\vspace{0.5cm}

\noindent {\bf Figure 2}. The $IW$ form factor $\xi(\omega)$ calculated 
for various values of the light-quark mass ratio $\eta = m_q / m_{q'}$. 
The value of  $\beta$ (Eq. (\ref{eq:beta})) is kept fixed at the value 
$\beta = 4$ and the $h.o.$ radial wave function (Eq. (\ref{eq:ho})) is 
considered. The solid, dashed, dotted and dot-dashed lines correspond to 
$\eta = 1.00, 0.52, 0.26$ and $0.13$, respectively.

\newpage

\begin{figure}[htb]

\centerline{\epsfxsize=12cm \epsfig{file=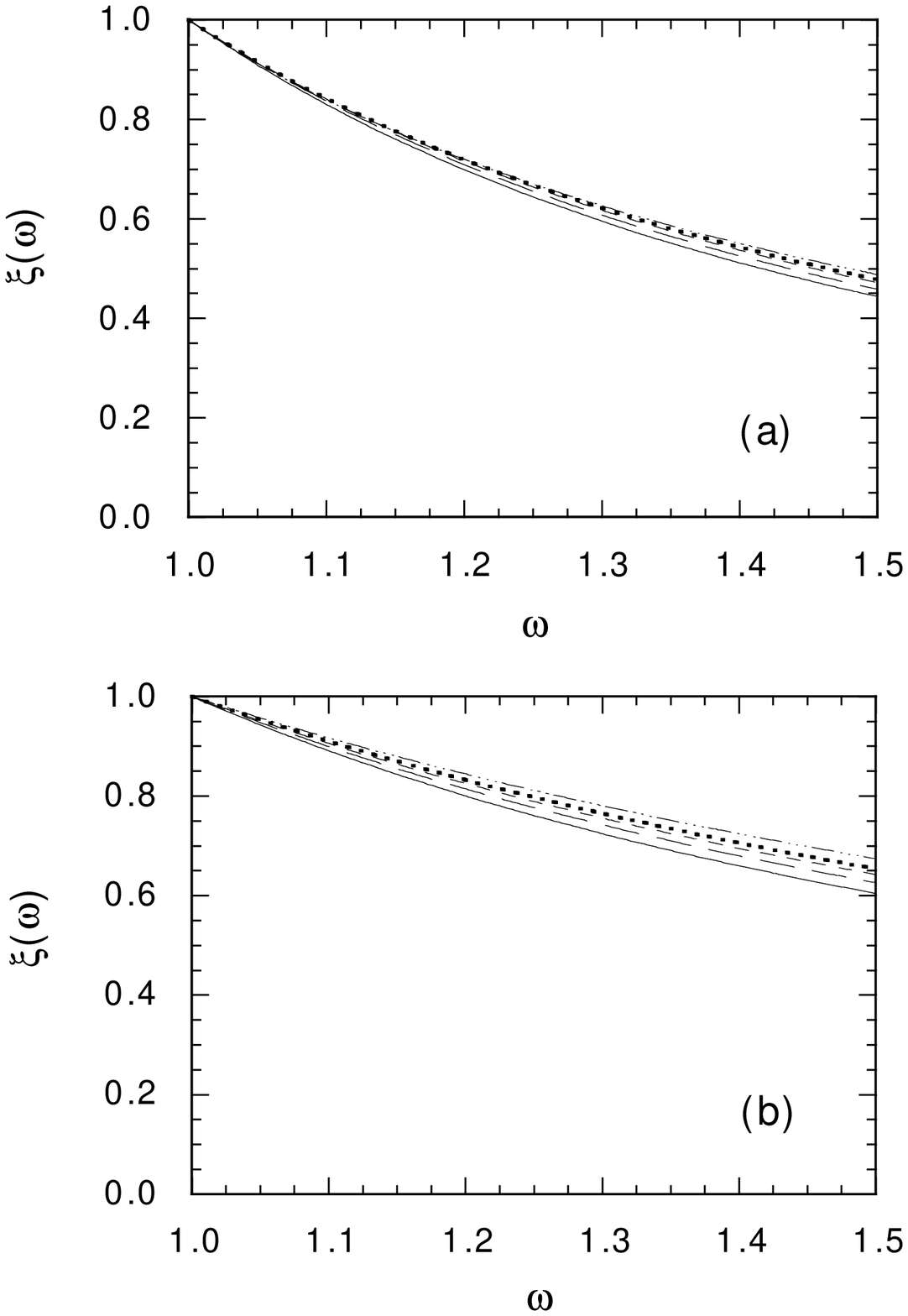}}

\end{figure}

\vspace{0.5cm}

\noindent {\bf Figure 3}. The $IW$ form factor $\xi(\omega)$ calculated 
at $\beta = 2$ (a) and $\beta = 10$ (b) using different baryon wave 
functions: $h.o.$ (solid lines); $p.l.$ with $n_p = n_k = 2, 3, 4$ and 
$8$ (dot-dashed, dotted, dashed and long-dashed lines, respectively).

\newpage

\begin{figure}[htb]

\centerline{\epsfxsize=11.5cm \epsfig{file=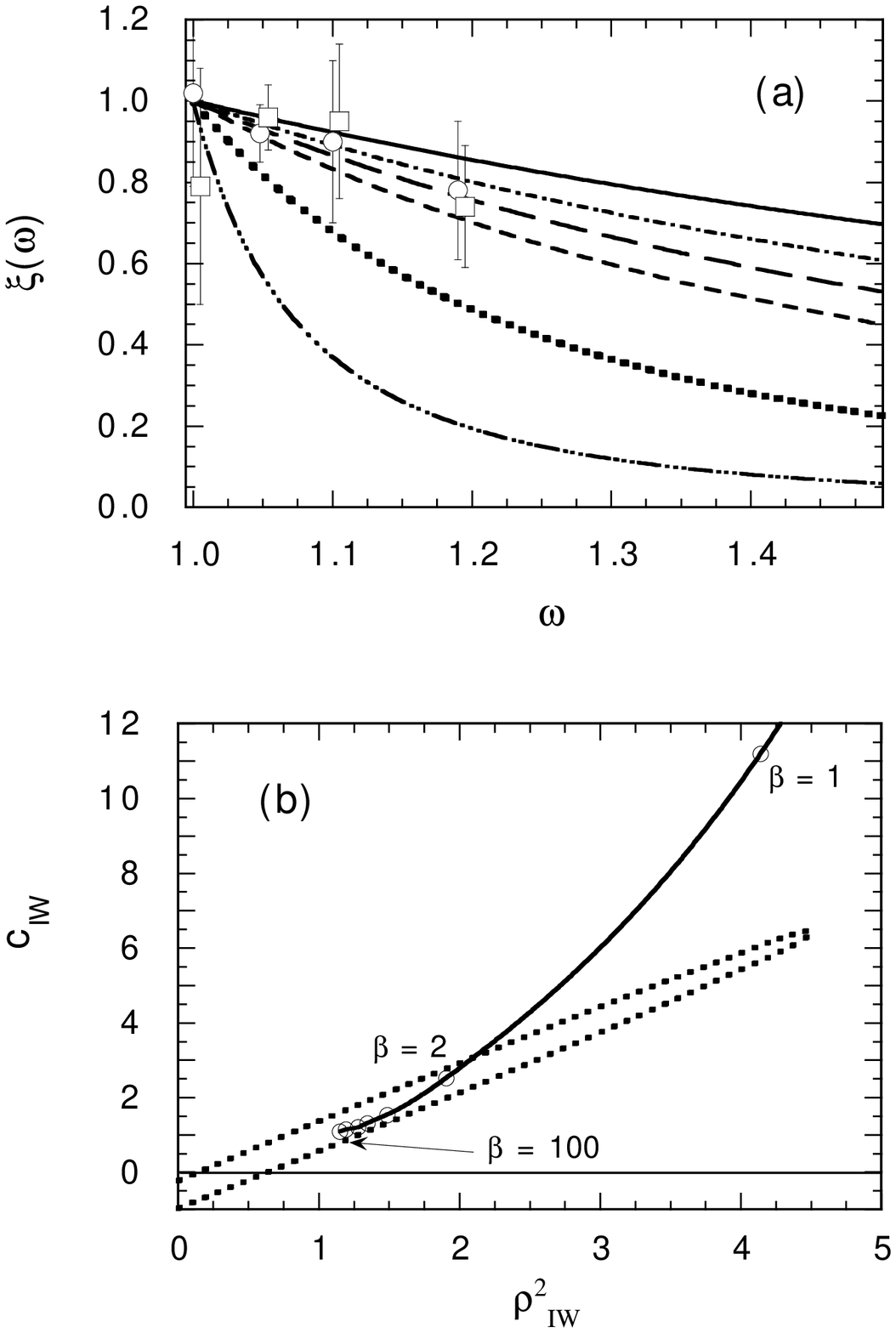}}

\end{figure}

\vspace{0.5cm}

\noindent {\bf Figure 4}. (a) The $IW$ function $\xi(\omega)$ calculated for
different values of $\beta$ and compared with the lattice $QCD$ calculations
of Ref. \cite{UKQCD} (open dots and squares). The triple-dotted-dashed,
dotted, dashed, long-dashed and dot-dashed lines correspond to $\beta = 0.5,
1.0, 2.0, 3.0$ and $10.0$, respectively, using the $h.o.$ wave function
(\ref{eq:ho}). The solid line is the result obtained at $\beta = 100$ using
the $p.l.$ wave function (\ref{eq:pl}) with $n_p = n_k = 2$. Adapted from
Ref. \cite{plb98}. (b) Curvature $c_{IW}$ versus slope $\rho_{IW}^2$ for the
$IW$ function $\xi(\omega)$. The dotted lines identify the allowed domain
determined in Ref. \cite{gupta}. The open dots correspond to our results
obtained for various values of $\beta$ and the solid line is just an
interpolation curve.

\newpage

\begin{figure}[htb]

\centerline{\epsfxsize=16cm \epsfig{file=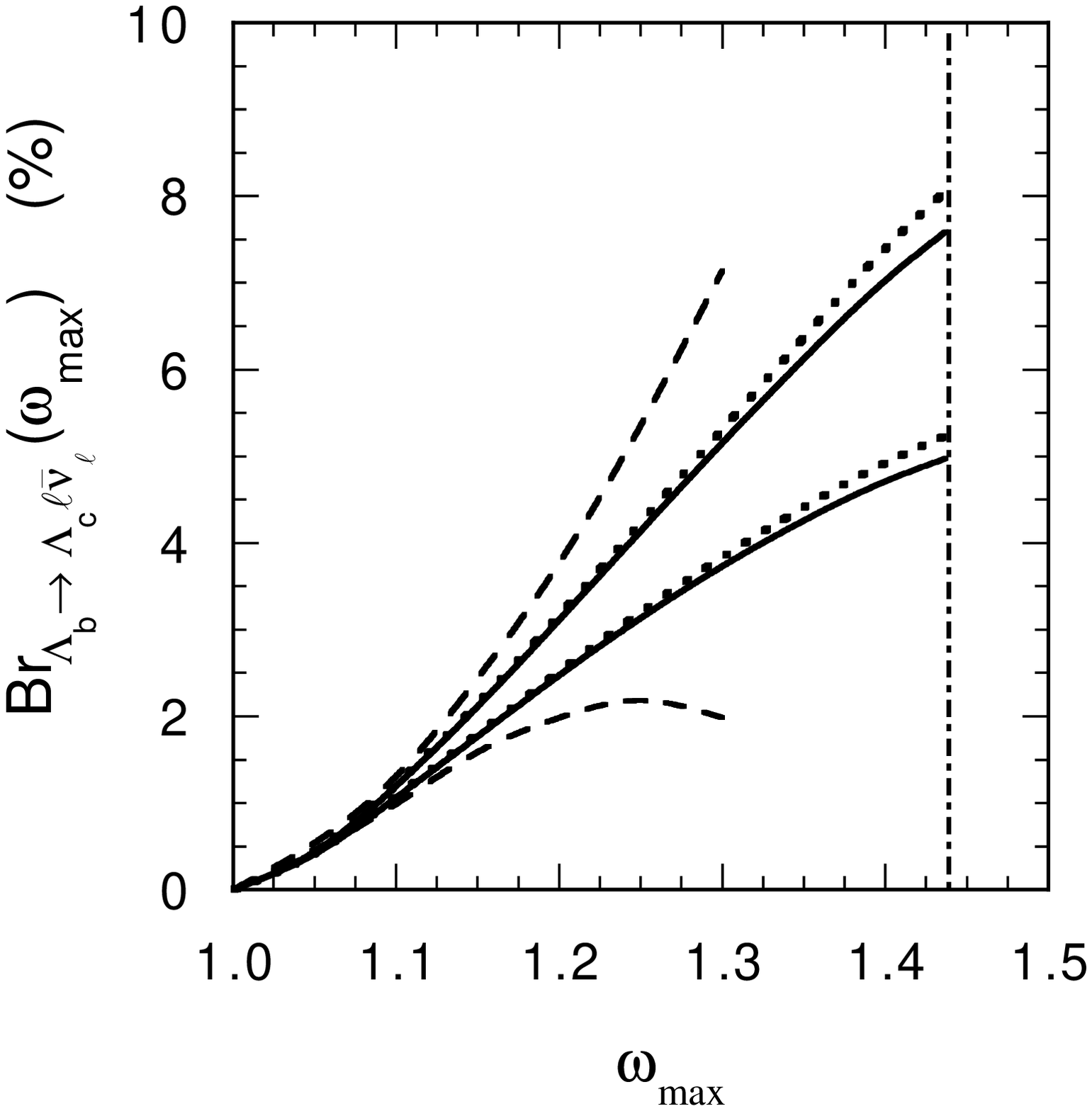}}

\end{figure}

\vspace{0.5cm}

\noindent {\bf Figure 5}. Partially integrated exclusive semileptonic 
branching ratio $Br_{\Lambda_b \to \Lambda_c \ell 
\bar{\nu}_{\ell}}(\omega_{max})$ in $\%$ versus $\omega_{max}$ (see Eq. 
(\ref{eq:Br})), calculated at $|V_{bc}| = 0.040$ and $\tau(\Lambda_b) = 
1.24 ~ ps$ \cite{PDG98} for the decay process $\Lambda_b \to \Lambda_c + 
\ell \bar{\nu}_{\ell}$. The solid and dashed lines correspond to our and 
lattice $QCD$ results \cite{UKQCD}, respectively. The lower and upper solid 
lines are the results corresponding to the use of $\xi(\omega) = 
\xi_L(\omega)$ and $\xi(\omega) = \xi_U(\omega)$ in Eq. 
(\ref{eq:first-o+pQCD}), respectively, i.e. including radiative plus 
first-order $1 / m_Q$ corrections with $\Lambda = 0.75 ~ GeV$. The dotted 
lines are the $HQS$ results (see Eqs. (\ref{eq:HQS})). The vertical 
dot-dashed line indicates the physical threshold $\omega_{th} \simeq 1.44$.

\newpage

\begin{figure}[htb]

\centerline{\epsfxsize=16cm \epsfig{file=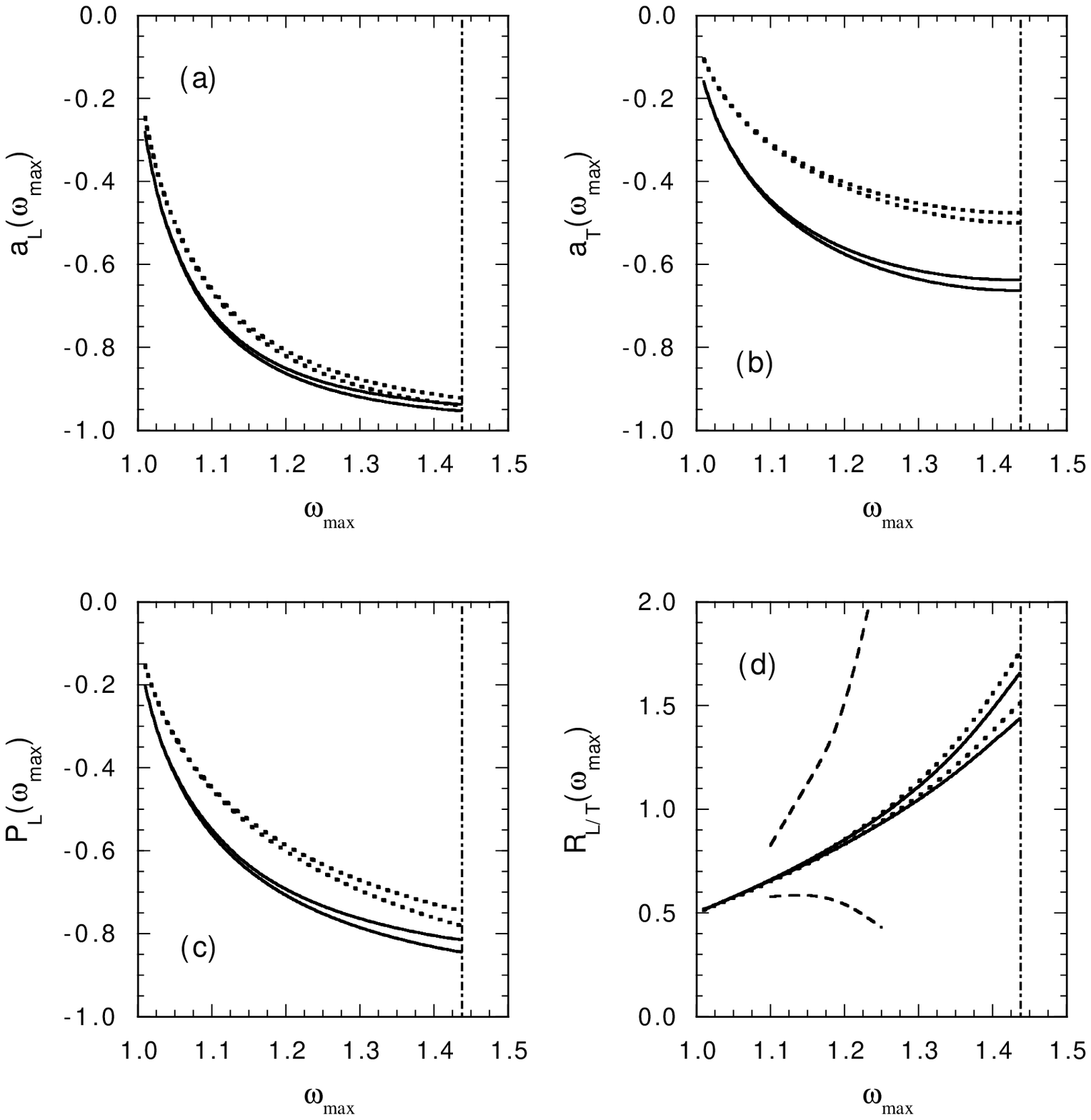}}

\end{figure}

\vspace{0.5cm}

\noindent {\bf Figure 6}. Partially integrated longitudinal asymmetry 
$a_L(\omega_{max})$ (a), transverse asymmetry $a_T(\omega_{max})$ (b), 
longitudinal daughter-baryon polarisation $P_L(\omega_{max})$ (c) and 
longitudinal to transverse decay ratio $R_{L/T}(\omega_{max})$ (d) versus 
$\omega_{max}$ for the decay process $\Lambda_b \to \Lambda_c + \ell 
\bar{\nu}_{\ell}$. The dotted and solid lines correspond to the results 
obtained in the heavy-quark limit (see Eqs. (\ref{eq:HQS})) and using Eqs. 
(\ref{eq:first-o+pQCD}) with $\Lambda = 0.75 ~ GeV$, respectively. The 
lower and upper lines are the results corresponding to $\xi(\omega) = 
\xi_L(\omega)$ and $\xi(\omega) = \xi_U(\omega)$, respectively. In (d) 
the dashed lines correspond to the lattice $QCD$ results of Ref. 
\cite{UKQCD}. The vertical dot-dashed lines indicate the physical 
threshold $\omega_{th} \simeq 1.44$.


\begin{thebibliography}{99}

\bibitem{iw} N. Isgur and M.B. Wise, Nucl. Phys. {\bf B348}, 276 (1991).

\bibitem{plb98} F. Cardarelli and S. Simula, Phys. Lett. {\bf B421}, 295 
 (1998) and in Proceedings of the III Int'l Conference on {\em Quark
 Confinement and the Hadron Structure}, Jefferson Lab (USA), June 1998, in
 press.

\bibitem{UKQCD} $UKQCD$ collaboration, K.C. Bowler et al., Phys. Rev. {\bf
 D57}, 6948 (1998).

\bibitem{bigi} For an updated analysis of the theoretical uncertainties 
 affecting the extraction of $V_{bc}$ see I.I. Bigi et al.: Annu. Rev. Nucl.
 Part. Sci. {\bf 47}, 591 (1997).

\bibitem{VEGAS} G.P. Lepage, J. Comp. Phys. {\bf  27}, 192 (1978).

\bibitem{gupta} D. Chakraverty, T. De, B. Dutta-Roy and K.S. Gupta, preprint
 SINP-TNP-98-04, e-print archive hep-ph/9802223, to appear in Int. J. Mod. 
 Phys. {\bf A}.

\bibitem{thr} C.E. Carlson et al.: Phys. Lett. {\bf B299}, 133 (1993). C.A. 
 Dominguez, J.K. K\"orner and D. Pirjol: Phys. Lett. {\bf B301}, 257 (1993). 
 B. Holdom and M. Sutherland: Z. Phys. {\bf C65}, 445 (1995).

\bibitem{PDG98} Particle Data Group, C. Caso et al., Eur. Phys. J. {\bf C3},
 1 (1998).

\bibitem{korner} J.G. Korner et al., Prog. Part. Nucl. Phys. {\bf 33}, 787
 (1994).

\bibitem{neubert} M. Neubert: Phys. Rept. {\bf 245}, 259 (1994).

\bibitem{QCDSR} Y. Dai, C. Huang, M. Huang and C. Liu: Phys. Lett. {\bf
 B387}, 379 (1996).

\bibitem{IMF} B. K\"onig, J.G. K\"orner, M. Kr\"amer and P. Kroll: Phys. 
 Rev. {\bf D56}, 4282 (1997).

\bibitem{korner_2} J.G. K\"orner and D. Pirjol, Phys. Lett. {\bf B334}, 399 
 (1994).

\bibitem{holdom_1} B. Holdom and M. Sutherland: Phys. Rev. {\bf D47}, 5067 
 (1993); Phys. Lett. {\bf B313}, 447 (1993); Phys. Rev. {\bf D48}, 5196 
 (1993).

\bibitem{holdom_2} B. Holdom, M. Sutherland and J. Mureika, Phys. Rev. {\bf 
 D49}, 2359 (1994).

\bibitem{UKQCD_M} $UKQCD$ Collaboration, K.C. Bowler et al.: Phys. Rev. {\bf 
 D54}, 3619 (1996).

\end{thebibliography}
\end{document}